\documentclass{appolb}
\usepackage{graphicx}
\usepackage{amssymb,bm}
\newcommand{\beq}[2]{\begin{equation}#1\label{#2}\end{equation}}
\newcommand{\ceq}[1]{(\ref{#1})}
\begin{document}
\author{Franco Ferrari \and
Jaros{\l}aw Paturej
\address{Institute of Physics and CASA*, University of Szczecin,
  ul. Wielkopolska 15, 70-451 Szczecin, Poland}}
\title{Diffusion of Brownian particles and Liouville field theory}
\maketitle

\begin{abstract}
In this work we review
a recently proposed
transformation which is useful
in order to
simplify non-polynomial
potentials  given in the form of an exponential.
As an application, it is shown that the
Liouville field theory may be mapped into a field theory with a
polynomial interaction
between
two scalar fields and
a  massive vector field.
The used methodology is illustrated with the help of the simple case
of a particle performing a random walk in a delta function potential.
\end{abstract}
\maketitle
\section{Introduction}\label{sec:intro}
The Liouville field theory (LFT) is traditionally related to string theory
and to  mathematical applications in the
complex geometry of Riemann surfaces. It has been for years a longstanding
problem whose solution has required outstanding efforts. It
would be impossible to cite in this short article
all the relevant works on this subject, so we
just give here a limited and very incomplete list of them
\cite{Pol3,Pol4,N,LAG,Ma1,Pol5,TA1,TA2,TZ1,Ma2,CMS,HJP,Tes1JLG,
jackiw1}. 
The intense research dedicated to LFT has greatly expanded our
knowledge of it. 
Recently, the so-called
three point correlation function of the theory has been explicitly
derived. All the other
correlation functions
can be computed using a recursive algorithm, so that 
the LFT has finally been proved to be solvable. 
One of the main difficulties in treating the LFT is the
fact that the potential of the theory consists of the exponential
$V_L(\phi)=\mu e^{-b\phi}$, where $\phi$ is the Liouville field and
$\mu,b$ are constants. This kind of nonpolynomial potentials
is not a problem in quantum mechanics or in the theory of the random
walk, where a limited number of particles are considered. It becomes
however awkward in a field theory, where there are infinite
degrees of freedom.
Despite this fact,
it is possible to compute the vacuum
expectation values of the observables of LFT using
a consistent perturbative approach \cite{jackiw1,jackiw2}.
The LFT has also applications in
statistical physics.
For instance, the grand canonical partition function 
of
a particle in a $2-$dimensional random potential whose correlation
functions grow logarithmically with the distance
coincides to the partition function of LFT \cite{carpledous}.
Moreover,
in \cite{Ma1, Mir1} 
the LFT has been associated to quantum
chaos in anyon systems. 
In \cite{Mu1} the identity $\frac{(-1)^N}{(N-1)!}\int d\mu\mu^{N-1}
e^{d^2x \mu e^{-b\phi}}=\frac 1{\left(
\int d^2xe^{-b\phi}
\right)^N}$ has been used to simplify the normalization of certain
vacuum expectation values.

Even if our present understanding of the LFT is
excellent, this model continues to inspire new ideas.
A fresh look at LFT comes this time from its connection to
random walks.  It turns out that the LFT may be ``linearized'',
i. e. it is possible to map the LFT with its nonpolynomial interaction
into another field theory, in which the interaction is polynomial.
More specifically, it has been shown that the
partition function of LFT is related to the grand canonical
partition function of a system of particles performing
a random walk in the presence of a disorder
vector field \cite{FePaLiouv}. In the equilibrium limit the two
partition functions coincide.
The strategy used to map the LFT into a statistical model
will be explained here starting from a simple example, namely that of
a random walk in a delta function potential.
Other physical examples will be discussed in the next
two Sections. Let us note that the opposite strategy, which consists
in rewriting the grand canonical partition function
of particles in the form of a field theory, is known since a long time
and it is now part of classical books on the subject of statistical
physics, see for instance the
case of particles in the Yukawa potential 
discussed in Ref.~\cite{zinnjustin}, Appendix A.27.

The presented material is divided as follows.
In Section~\ref{bbb} we discuss the case of a particle performing a random
walk in the presence of a delta function potential. In
Section~\ref{secc} 
two cases are presented, in which thanks to the
introduction 
of auxiliary fields  it has been possible to simplify complicated
interactions, 
allowing in this way
to achieve a better knowledge of polymer systems.
Section~\ref{secd} is dedicated to the theory of Liouville.
Finally our conclusions are presented in Section~\ref{sece}.

\section{Random walk in a delta function potential}\label{bbb}
We present in this Section a simple case, which will
be used as a toy model
 to illustrate the strategy for
``linearizing'' the LFT.

Let us consider the random walk of a particle in $d-$dimensions around
an infinitely narrow and deep potential well located at the point
$\mathbf r_0$. This kind of obstacles is described by the
Dirac delta function
potential $V(\mathbf r(t))=-v_0\delta(\mathbf r(t)-\mathbf r_0)$,
$v_0>0$, where  $\mathbf r(t)$ is the radius vector which denotes the
position of the particle at each instant $t$. According to the duality
between statistical mechanics and quantum mechanics, the partition
function $Z_\delta(T;\mathbf r_2,\mathbf r_1)$ of our particle may be formally obtained
from the partition function of a quantum particle immersed in the
repulsive potential 
$+v_0\delta(\mathbf r(t)-\mathbf r_0)$  after continuing the time to
imaginary values:
\beq{Z_\delta(T;\mathbf r_2,\mathbf r_1)=\int_{\mathbf r(0)=\mathbf r_1}^{\mathbf r(T)=\mathbf
    r_2}
{\cal D}\mathbf r(t)\exp\left\{
-\int_0^Tdt\left(
\dot{\mathbf r}^2(t)+v_0\delta(\mathbf r(t)-\mathbf r_0)
\right)
\right\}
}
{zdelta}
In the above equation $\mathbf r_1$ and $\mathbf r_2$ are the
positions of the particle at the initial and final instants $t=0$ and
$t=T$ respectively.
It turns out that, after a Laplace transform with respect to the time
$T$, $Z_\delta(T;\mathbf r_2,\mathbf r_1)$ satisfies a
Schr\"odinger--like equation 
which is solvable.

We will now concentrate on the delta function potential. It is not
difficult to convince ourselves that this is a complicated, nonlinear
potential. To this purpose, it is sufficient to write it using the
Fourier representation:
\beq{
\delta(\mathbf r(t)-\mathbf r_0)=\int d^d ke^{-i\mathbf k\cdot(\mathbf
  r(t)-\mathbf r_0)}
}{fourep}
It is possible to linearize this potential at the price of introducing
fields using the following identity:
\beq{
e^{
-\int_0^Tdt v_0\delta(\mathbf r(t)-\mathbf r_0)}=
\int{\cal D}\varphi_1(\mathbf x){\cal D}\varphi_2(\mathbf x)
e^{-i\int d^d\mathbf x\varphi_1\left[
\frac 1{v_0}\varphi_2-J_1 
\right]
}e^{-\int d^d\mathbf x \varphi_2J_2}
}{idesimp}
where
\beq{
J_1(\mathbf x)=\int_0^Tdt\delta(\mathbf x-\mathbf
r(t))\qquad\mbox{and}\qquad J_2(\mathbf x)=\delta(\mathbf x-\mathbf r_0)
}{currspec}
To prove Eq.~\ceq{idesimp}, we
integrate over the field $\varphi_1$ in the right hand side of
Eq.~\ceq{idesimp}:
\beq{\int{\cal D}\varphi_1{\cal D}\varphi_2
e^{
-i\int d^d\mathbf x\varphi_1\left[
\frac 1{v_0}\varphi_2-J_1
\right]
}e^{-\int d^d\mathbf x \varphi_2J_2}
=\int{\cal D}\varphi_2\delta\left({\textstyle
\frac{\varphi_2}{v_0}-J_1}
\right)e^{-\int d^d\mathbf x \varphi_2J_2}
}{proofone}
In other words, the field $\varphi_1(\mathbf
x)$ in Eq.~\ceq{idesimp} 
is just a Lagrange multiplier enforcing the condition
$\varphi_2=v_0J_1$.
A simple integration over the remaining field $\varphi_2$ in
Eq.~\ceq{proofone} gives as a result:
\beq{\int{\cal D}\varphi_2(\mathbf x)
\delta\left({\textstyle
\frac{\varphi_2}{v_0}-J_1}
\right)e^{-\int d^d\mathbf x \varphi_2J_2}=e^{-\int d^d\mathbf x
  v_0J_1(\mathbf x)J_2(\mathbf x)}
}{prooftwo}
The proof of identity \ceq{idesimp} is completed by noticing that
the external currents $J_1,J_2$ defined in Eq.~\ceq{currspec} have
been chosen exactly in such a way that:
$\int d^d\mathbf x
  v_0J_1(\mathbf x)J_2(\mathbf x)=v_0\int_0^Tdt\delta(\mathbf
  r(t)-\mathbf r_0)$.

Using Eq.~\ceq{idesimp}, one may write the partition function
$Z_\delta(T;\mathbf r_2,\mathbf r_1)$ 
as follows:
\begin{eqnarray}Z_\delta(T;\mathbf r_2,\mathbf r_1)
&=&\int_{\mathbf r(0)=\mathbf r_1}^{\mathbf r(T)=\mathbf r_2}{\cal
  D}\mathbf r(t)
{\cal D}\varphi_1(\mathbf x){\cal D}\varphi_2(\mathbf x)
e^{-\int_0^Tdt\left[\dot{\mathbf r}^2(t)+i\varphi_1(\mathbf
  r(t))\right]}\nonumber\\
&\times&
e^{-i\int d^d\mathbf x\varphi_1\frac 1{v_0}\varphi_2}e^{-\int
  d^d\mathbf x \varphi_2J_2}
\label{zdafteridesimp}
\end{eqnarray}
As it stands, $Z_\delta(T;\mathbf r_2,\mathbf r_1)$  describes a
random walk in the imaginary random potential $i\varphi_1(\mathbf
r(t))$.
A more physical interpretation of $Z_\delta(T;\mathbf r_2,\mathbf
r_1)$ may be achieved by considering its Laplace transform with
respect to $T$:
\beq{
Z_\delta(E;\mathbf r_2,\mathbf r_1)=
\int_0^{+\infty}dTe^{-TE}Z_\delta(T;\mathbf r_2,\mathbf r_1)
}{lapltransf}
The variable $E$ plays the same role of the energy in quantum
mechanics. For this reason, it will be called pseudo--energy.
It is convenient to rewrite the expression of
$Z_\delta(E;\mathbf r_2,\mathbf r_1)$  as follows:
\beq{Z_\delta(E;\mathbf r_2,\mathbf r_1)
=\int{\cal D}\varphi_1\int{\cal D}\varphi_2
\Xi(E;\mathbf r_2,\mathbf r_1,[\varphi_1])
e^{-i\int d^d\mathbf x\varphi_1\frac 1{v_0}\varphi_2}e^{-\int
  d^d\mathbf x \varphi_2J_2}
}{splitzd}
where
\begin{eqnarray}
\!\!\!\!\!\!\!\!\!\!
\Xi(E;\mathbf r_2,\mathbf r_1,[\varphi_1])
&=&\int_0^{+\infty}dTe^{-TE}
\int_{\mathbf r(0)=\mathbf r_1}^{\mathbf r(T)=\mathbf r_2}{\cal
  D}\mathbf r(t)e^{-\int_0^Tdt\left[\dot{\mathbf r}^2(t)+i\varphi_1(\mathbf
  r(t))\right]}
\label{xifundef}\end{eqnarray}
Now we apply to $\Xi(E;\mathbf r_2,\mathbf r_1,[\varphi_1])$
the formula:
\begin{eqnarray}
\!\!\!\!\!\!\!\!\!\!
\Xi(E;\mathbf r_2,\mathbf r_1,[\varphi_1])
&=&
\lim_{n\to 0}\int{\cal D}\vec{\phi}(\mathbf
x)\phi_1(\mathbf r_2)\phi_1(\mathbf r_1)e^{-\frac 12\int d^2\mathbf x
\left[
(\mathbf\nabla\vec\phi)^2+E\vec\phi^2+i\varphi_1\vec\phi^2
\right]
}
\label{formreptri}\end{eqnarray}
The above equation is a variant of the so--called replica trick that
can be found in classical textbooks \cite{zinnjustin,kleinert}.
The replica field $\vec\phi$ is a $n-$components vector field
$\vec\phi=(\phi_1,\ldots,\phi_n)$.
The limit $n\longrightarrow 0$ in Eq.~(\ref{formreptri}) has the
following meaning. The path integral in the right hand side should be
computed for integer values of $n$ such that $n\ge 1$.
In this way, the field theory of Eq.~(\ref{formreptri}) becomes a
$O(n)$ scalar field theory and its correlation functions may be
derived by standard techniques \cite{kleinert2}. The result of such
calculations is then continued analytically to any value of $n$.
At the end, one should take the limit $n\longrightarrow0$. Such a
limit is 
nontrivial. It is possible to show for instance that certain
discrete spin
systems with 
$O(n)$ symmetry describe a self--avoiding random walk  in the limit
$n\longrightarrow 0$ \cite{cardy,vilgisrep}.
In the continuous  case it is possible to see that many Feynman
diagrams vanish in a $O(n)$ field theory when $n$ approaches zero, but
even in the one-loop approximation some of them are able to survive
and deliver a finite expression of the partition function of polymer
systems \cite{FeLaNPB, necros}. 

Substituting Eq.~\ceq{formreptri} in
Eq.~\ceq{splitzd} the partition function $Z_\delta(E;\mathbf
r_2,\mathbf r_1)$ becomes:
\begin{eqnarray}
Z_\delta(E;\mathbf r_2,\mathbf r_1)&=&\lim_{n\to0}\int{\cal D}\vec\phi
{\cal D}\varphi_1 {\cal D}\varphi_2e^{-\frac 12
\int d^d\mathbf x\left[
(\mathbf\nabla\vec\phi)^2+E\vec\phi^2+i\varphi_1\vec\phi^2
\right]}\nonumber\\
&\times&
e^{-i\int d^d\mathbf x\varphi_1\frac 1{v_0}\varphi_2}e^{-\int
  d^d\mathbf x \varphi_2J_2}\phi_1(\mathbf
r_1)\phi_1(\mathbf r_2)
\label{zdaftreptri}
\end{eqnarray}
A straightforward integration over the fields $\varphi_1,\varphi_2$
produces the result:
\beq{
Z_\delta(E;\mathbf r_2,\mathbf r_1)=
\lim_{n\to 0}
\int{\cal D}\vec\phi(\mathbf x)e^{-\frac 12\int d^d\mathbf x\left[
(\mathbf\nabla \vec\phi)^2+E\vec\phi^2
\right]}e^{\frac {v_0}2\vec\phi^2(\mathbf r_0)}\phi_1(\mathbf
r_1)\phi_1(\mathbf r_2)
}{zdfin}
Concluding, the partition function $Z_\delta(E;\mathbf r_2,\mathbf
r_1)$ 
of the initial
random walk has been
transformed in a path integral describing a gaussian field theory. 
It is easy to realize after taking
the limit $n\longrightarrow 0$ that
$Z_\delta(E;\mathbf r_2,\mathbf
r_1)$ coincides with the propagator of a scalar field theory with
action
$S=\frac 12\int d^d\mathbf x\left[
(\mathbf\nabla \phi_1)^2+
E\phi_1^2-v_0\delta(\mathbf x-\mathbf
r_0)\phi_1^2(\mathbf x)\right] $.
From this action it is possible to derive the differential
equation satisfied
by $Z_\delta(E;\mathbf r_2,\mathbf r_1)$:
\beq{
\left[
E-\Delta_{\mathbf r_2}-v_0\delta(\mathbf r_2-\mathbf r_0)
\right]Z_\delta(E;\mathbf r_2,\mathbf r_1)=\delta(\mathbf r_2-\mathbf r_1)
}{pseudschroed}
$\Delta_{\mathbf r_2}$ is the  Laplacian with respect to the
coordinates $\mathbf r_2$. $Z_\delta(E;\mathbf r_2,\mathbf r_1)$
satisfies also an analogous equation with $\mathbf r_2$ replaced by
$\mathbf r_1$.
Eq.~\ceq{pseudschroed} is exactly the
solvable Schr\"odinger--like equation mentioned after
Eq.~\ceq{zdelta} that 
appears also in other  problems in which particles move in a
delta function potential.
For completeness, we give the solution of
Eq.~\ceq{pseudschroed} \cite{schulman}:
\beq{
Z_\delta(E;\mathbf r_2,\mathbf r_1)=Z_0(E;\mathbf r_2,\mathbf r_1)+
\frac{Z_0(E;\mathbf r_2,\mathbf r_0)Z_0(E;\mathbf r_0,\mathbf
  r_1)}{Z_0(E;\mathbf r_0,\mathbf r_0)-\frac 1{v_0} }
}{formsol}
where $Z_0(E;\mathbf r_2,\mathbf r_1)$ is the Green function
satisfying the free equation 
\beq{
\left[E-\Delta_{\mathbf
    r_2}\right]Z_0(E;\mathbf r_2,\mathbf r_1) =\delta(\mathbf
r_2-\mathbf r_1)}{freeeq}
We remark that, as it stands, the above solution is just formal,
because
the Green function
$Z_0(E;\mathbf r_2,\mathbf r_1)$ computed at coinciding points $\mathbf
r_2=\mathbf r_1=\mathbf r_0$ is singular. To eliminate this
singularity a suitable prescription is necessary, see for example
\cite{Grosche} for more information on this subject. 

\section{The case of polymer physics }\label{secc}
A
random flight chain composed by $N$ segments of equal length $a$
when $N$ becomes large and $a$ becomes small looks like the trajectory
of a fluctuating particle whose average free path amounts to $a$.
The random flight chain is the basic model of a polymer. For this reason,
polymer physics and the theory of random
walks are related. Of course, the trajectory of a
particle may intersect itself many times, but this is
not the case of a polymer chain due to the strong repulsions between
the electron layers of the molecules composing the chain. 
These repulsive interactions have been modeled by
de Gennes and coworkers with the help of the 
so--called self-avoiding potential. We give here its expression for a
continuous chain:
\beq{
V(1)=\int_0^Lds\int_0^L
ds'\frac{v_0}{2}\delta^{(3)}(\mathbf r(s)-\mathbf r(s'))}
{repint}
In writing the above equation
the chain, which is supposed to be without branches,
 has been parametrized using its arc--length $s$.
The initial and final points of the random walk $r(0)=\mathbf r_1$ and
$\mathbf r(T)=\mathbf r_2$ have been replaced by the ends of the
polymers $\mathbf 
r(0)$ and $\mathbf r(L)$. $L$ is the total length of the chain.

As in the previous case, also the self-avoiding potential contains a
delta function, which may be 
``linearized'' 
applying a variant of the Hubbard--Stratonovich
transformations:
\beq{
e^{-V(1)}=\int{\cal D}\varphi e^{-A_\varphi(1)}
\exp\left\{
-i\int_0^Lds\varphi(\mathbf r(s))
\right\}
}{hubstra}
where the action of the $\varphi$ field is given by
$A_\varphi(1)=\frac{1}{2v_0}\int d^3 x\varphi^2(\mathbf x)$.

Another system with
a complicated potential which needs simplification to make the theory
tractable is that of two closed chains $A$ and $B$
of lengths $L_A$ and $L_B$ respectively subjected to
topological constraints.
The simplest topological invariant that can be used to impose
topological constraints in three dimensions
is the Gauss linking number $\chi(A,B)$. For two closed curves
$\mathbf r_A(s)$ 
and $\mathbf r_B(s)$ representing the polymer chains $\chi(A,B)$
can be expressed as follows:
\beq{
\chi(A,B)=\int_0^{L_1}ds_A\dot{\mathbf r}_A(s_A)\cdot{\mathbf
  b}_{B}(\mathbf r_A(s_A)) 
}{galinv}
Here ${\mathbf b}_B$ is the magnetic field generated by the
fictitious current ${\mathbf j}_B(\mathbf x)=\frac{1}{4\pi}
\int_0^{L_B}ds_B\dot{\mathbf r}_B(s_b)\delta(\mathbf x-\mathbf
r_B(s_B))$ flowing along the chain $B$:
\beq{
{\mathbf b}_B(\mathbf x)=
\frac {1}{4\pi}
\int_0^{L_B}ds_B\dot{\mathbf r}_B(s_B)
\times\frac{(\mathbf x-\mathbf r_B(s_B))}{|\mathbf x
  -\mathbf r_B(s_B)|}
}{magfiedef}
It is possible to check that $\mathbf \nabla\times{\mathbf
  b}_B=\mathbf j_B $ and that $\mathbf \nabla\cdot\mathbf b_B=0$,
justifying the interpretation of $\mathbf b_B$ as a magnetic field.

In the action which describes the fluctuations of the two chains $A$
and $B$, the effect of
the topological conditions  imposed on their trajectories is the
appearance of the fictitious magnetic interaction term given in equation
\ceq{galinv}.
The analog of the identities \ceq{idesimp} and \ceq{hubstra}, which
simplifies in the present case the 
interaction of topological origin and
allows to
 extract from the theory physical predictions \cite{FeAnn},
is given by:
\beq{
e^{-i\chi(A,B)}=
\int {\cal D}\mathbf A(\mathbf x)\mathbf B(\mathbf x)
e^{-iA_{CS}+\frac{1}{4\pi}\int d^3\mathbf x\mathbf J_A\cdot
  \mathbf A
+\int d^3\mathbf x\mathbf J_B\cdot   \mathbf B
}
}{idetop}
where $A_{CS}$ is the action of the Chern--Simons fields $\mathbf A$
and $\mathbf B$:
\beq{
A_{CS}=i\int d^3\mathbf x \mathbf A\cdot(\mathbf \nabla \times\mathbf B)
}{csaction}
and
\beq{
\mathbf J_A(\mathbf x)=\int_0^{L_A}ds_A\delta(\mathbf x-\mathbf
r_A(s_A))\qquad
\mathbf J_B(\mathbf x)=\int_0^{L_B}ds_B\delta(\mathbf x-\mathbf r_B(s_B))
}{topcurr}
\section{The Liouville field theory }\label{secd}
In this Section we consider the Liouville action on the Euclidean plane:
\beq{
Z_L=\int{\cal D}\phi e^{-\int d^2x\left[ \frac 12 (\mathbf
    \nabla\phi)^2+\mu e^{-b(\phi(\mathbf x)-\phi(\mathbf r_0))}\right]}
}{liouvact}
Let's note that in the standard LFT the exponential
potential is given by $e^{-b\phi(\mathbf x)}$. 
To simplify the nonpolynomial interaction appearing in Eq.~\ceq{liouvact}
it is possible to use the identity:
\beq{
 e^{-\int d^d\mathbf x\mu e^{
 -b(\phi(\mathbf x)-\phi(\mathbf r_0))}}=
\lim_{T\to+\infty} \Xi(T,[\phi])
}{idecomplone}
where
\begin{eqnarray}
\Xi(T,[\phi])&=&\int{\cal D}\varphi_1(t,\mathbf x)
\int{\cal D}\varphi_2(t,\mathbf x) e^{-i\int
  d^d\mathbf x dt \left[\varphi_1\left(
\frac{\partial}{\partial t}-g(\mathbf \nabla+b\nabla\phi)^2
\right)\varphi_2\right]}\nonumber\\
&\times &e^{-i\int d^d\mathbf xdt\varphi_1J_1}e^{-\int d^d\mathbf x dt\varphi_2J_2}
\label{idecompltwo}
\end{eqnarray}
and
\beq{
J_1(t,\mathbf x)=\delta(\mathbf x-\mathbf r_0)\delta(t)\qquad\qquad
J_2(t,\mathbf x)=(4\pi gT)^{\frac d2}\mu\delta(T-t)
}{speccurcho}
This is actually a general identity which is valid in
$d-$dimensions. The left hand side of Eq.~\ceq{idecomplone} coincides
with the Liouville potential term when $d=2$.

Let us prove Eq.~\ceq{idecomplone}. Once again $\varphi_1$ is a
Lagrange multiplier that can be easily integrated out giving as a
result:
\beq{
\Xi(T,[\phi])=\int{\cal D}\varphi_2\delta\left[
\left(\frac{\partial}{\partial t}-g(\mathbf\nabla+b\mathbf
\nabla\phi)^2\right)\varphi_2-J_1
\right]e^{-\int d^d\mathbf x dt\varphi_2J_2}
}{xifdf}
The functional delta function in Eq.~\ceq{xifdf} forces the condition:
\beq{
\left[
\frac{\partial}{\partial t}-g(\nabla+b\mathbf\nabla\phi)^2
\right]\varphi_2=J_1
}{diffproc}
To derive the explicit expression of $\varphi_2$ we compute the
Green function of the associated point source equation:
\beq{
\left[
\frac{\partial}{\partial t}-g(\nabla+b\mathbf\nabla\phi)^2
\right]G(t-t';\mathbf x,\mathbf x',[\phi])
=\delta(t-t')\delta(\mathbf
x-\mathbf x')
}{diffprocgrefun}
The solution of Eq.~\ceq{diffprocgrefun} is: 
\beq{
G(t-t';\mathbf x,\mathbf x',[\phi])=e^{-b(\phi(\mathbf x)-\phi(\mathbf
  x'))}
\frac{\theta(t-t')}{(4\pi(t-t')g)^{\frac d2}}e^{-\frac{(\mathbf
    x-\mathbf x')^2}{4(t-t')g}}
}{gtxxp}
$\theta(t)$ is the
Heaviside theta function.
$G(t-t';\mathbf x,\mathbf x',[\phi])$
is the Green function of a stochastic process describing the
diffusion of a particle in a random potential $\mathbf\nabla \phi$.
Knowing $G(t-t';\mathbf x,\mathbf x',[\phi])$, it is possible to solve
Eq.~\ceq{diffproc} as follows:
\beq{
\varphi_2(t,\mathbf x)=\int d^d\mathbf x' dt'
G(t-t';\mathbf x,\mathbf x',[\phi])J_1(t',\mathbf x')
}{phitwo}
As a consequence, after the elimination of the remaining field $\varphi_2$
in Eq.~\ceq{xifdf}, we obtain:
\beq{ 
\Xi(T,[\phi])=e^{-\int d^d\mathbf x dt d^d\mathbf x' dt' J_2(t,\mathbf
  x)G(t-t';\mathbf x,\mathbf x',[\phi])J_1(t',\mathbf x') }
}{xifdhg}
Recalling the expression of the Green function $G(t-t';\mathbf
x,\mathbf x',[\phi])$ given in Eq.~\ceq{gtxxp} and
the special choice of currents \ceq{speccurcho}, we
get:
\beq{
\Xi(T,[\phi])=
\exp\left\{-\int d^d\mathbf x \enskip e^{-b (\phi(\mathbf
    x)-\phi(\mathbf x'))} e^{-\frac{(\mathbf x-\mathbf x')^2}{4Tg}}
\right\}}{profdf}
In the limit $T\longrightarrow +\infty$ the functional
$\Xi(T,[\phi])$ becomes 
exactly equal to the left hand side of Eq.~\ceq{idecomplone}.
This completes our proof.

Applying the identity in Eq.~\ceq{idecomplone} for $d=2$ in order to
simplify the potential of the Liouville action,
the partition function of the 
LFT may be
written as follows:
\begin{eqnarray}
Z_L&=&\lim_{T\to +\infty}\int{\cal D}\phi(\mathbf x){\cal
  D}\varphi_1(t,\mathbf x)
{\cal D}\varphi_2(t,\mathbf x) e^{-\frac 12 d^2\mathbf 
  x(\nabla\phi)^2}\nonumber\\
 && e^{-i\int
  d^2\mathbf x dt\left[\varphi_1\left(
\frac{\partial}{\partial t}-g(\mathbf \nabla+b\nabla\phi)^2
\right)\varphi_2\right]}e^{-i\int d^d\mathbf xdt\varphi_1J_1}e^{-\int
  d^d\mathbf xdt
\varphi_2J_2}
\label{liouvfin}
\end{eqnarray}
This is our final result. It is possible to note that the above field
theory has only polynomial 
interactions and it is very similar to scalar electrodynamics.

\section{Conclusions}\label{sece}
In Section \ref{bbb} we have  discussed in details
 the simple example of the random
walk of a particle in a delta function potential.
In Eq.~\ceq{zdfin} the
partition function $Z_\delta(E;\mathbf
r_2,\mathbf r_1)$ of this system
has been rewritten  in the form of
a gaussian field theory using the identity (\ref{idesimp}).
At this point, to derive
$Z_\delta(E;\mathbf
r_2,\mathbf r_1)$  it is sufficient to solve the pseudo--Schr\"odinger
equation \ceq{pseudschroed}. This result is in agreement with previous
research on the subject, see for instance \cite{Grosche}.

In Section~\ref{secc} we have presented two other identities,
Eqs.~\ceq{hubstra} and \ceq{idetop}, in which highly nonlinear
potentials have been 
transformed in polynomial interactions. The application of the identity
of Eq.~\ceq{hubstra} together with the replica trick  has brought 
astonishing advances in our understanding of polymer solutions and melts.
The advantages of Eq.~\ceq{idetop} in the case of
polymer systems with 
 topological constraints have been illustrated in Ref.~\cite{FeAnn}.

Finally, Section~\ref{sece} is dedicated to LFT. Eq.~\ceq{idecomplone}
is the generalization to $d-$dimensions 
of the two dimensional formula of Ref.~\cite{FePaLiouv}.
Using the identity \ceq{idecomplone}, it is possible to rewrite the LFT
as a field theory with polynomial interactions which closely resembles
scalar electrodynamics.

It is interesting to note the analogies and differences
 of our work on LFT with that of
Ref.~\cite{carpledous}. In \cite{carpledous} the statistical mechanics of
a particle moving in a gaussian scalar random potential $\phi$ has
been studied. 
If the two point function of this random potential grows logarithmically with
the distance, 
the grand canonical partition function of
the system coincides with LFT. 
Eq.~\ceq{liouvfin} can be regarded instead
as the average over the vector random 
potential $\nabla\phi$ of the functional
$\Xi(T,[\phi])$. As it has been shown in Ref.~\cite{FePaLiouv},
$\Xi(T,[\phi])$ is the grand canonical partition function 
 of a
system of particles 
performing a random walk starting from the point $\mathbf r_0$. 
It is quite remarkable that, in the  limit  $T\longrightarrow
+\infty$, the LFT is recovered. In other words,
the grand canonical partition function of Ref.~\cite{carpledous}
represents the equilibrium limit of $\Xi(T,[\phi])$.

The aim of this article and of \cite{FePaLiouv} was mainly to establish
the identity (\ref{idecomplone}). This
identity has also the effect of ``linearizing'' the
exponential potential of LFT. In this way, it becomes possible  to
exploit the 
standard techniques of field theory in the study of the Liouville
model.
Of course, as already mentioned in the Introduction, the LFT
 has been already solved with the method of conformal bootstrap.
Moreover, a perturbative approach to LFT has been
developed in \cite{jackiw1,jackiw2}.
Despite that, the form of LFT obtained here
makes it possible to use other techniques, such as for instance the loop
expansion successfully applied to scalar electrodynamics
\cite{colwei}.  
What is even more important, the present approach is open to several
extensions.  While  the method of
conformal 
bootstrap is limited to two dimensions, here we are able  to map
in principle also 
the $d-$dimensional Liouville model into a polynomial vector field
theory which the generalization
of Eq.~(\ref{liouvfin}) to $d-$dimensions. Of course, one should be
careful
in doing that,
because the Liouville field theory is nonrenormalizable in more than
two dimensions. 
Moreover, there are also other field theories which are not exactly
solvable, but have potentials of the exponential form. In that case
our technique may provide an useful tool to investigate such theories.
\section{Acknowledgements}
This work has been partially financed
by the Polish Ministry of Science and Higher Education, scientific
project N202~156~31/2933.  
F. Ferrari gratefully acknowledges also the support of the action
COST~P12 financed by the European Union and the hospitality of
C. Schick at the University of Rostock.
Both authors wish to thank the anonymous referee for helpful
suggestions.
They would also thank the Organizers for organizing such a beautiful
and fruitful Symposium.

\end{document}